# First-principles Calculations of High Thermal Conductivity in Germanium Carbide Channel Materials

S.-C. Lee[1], Y.-T. Chen[1], C.-R. Liu[1], S.-M. Wang[1], Y.-T. Tang[1*], F.-S. Chang[2], Z.-X. Li[2], K.-Y. Hsiang[2,3], M.-H. Lee[2]

1   Department of Electrical Engineering, National Central University, Taoyuan 320317, Taiwan
2   Institute of Electro-Optical Engineering, National Taiwan Normal University, Taipei 11677, Taiwan
3   Institute of Electronics, National Yang Ming Chiao Tung University, Hsinchu 300, Taiwan

*Abstract*

Silicon carbide (SiC) has become a popular material for next-generation power components due to its smaller size, faster switching speed, simpler cooling and greater reliability than Si-MOSFETs. With this in mind, we are thinking about whether the replacement of Si-base with Germanium Carbide(GeC) will also have good performance. This work explains the heat transfer of GeC by simulating the thermal conductivity through molecular dynamics (MD) and proposes a potential 4H-GeC power MOSFET with wide bandgap and high thermal conductivity to replace Si-MOSFETs

*Introduction*

When looking for the next generation of high-efficiency power converter switches, wide-bandgap semiconductor materials such as gallium nitride (GaN) and silicon carbide (SiC) are the ideal choices. They both have band gaps wider than 3 eV. Compared with silicon, wide bandgap semiconductor devices bring significant power efficiency to a variety of applications, for example, automotive and battery chargers, PV converters, server power supplies, and circuit protection. Their application has gradually replaced silicon and changed our world. Silicon carbide is an excellent semiconductor for power applications that allows 10-times higher critical field strength in channels due to the wide band gap. For this, SiC chips can be made thinner with lower power losses. With strong tetrahedral covalent bonds, SiC exhibits excellent heat stability. Thermal conductivity is an indicator that a semiconductor dissipates the generated heat. With a thermal conductivity 10-times higher than silicon and alumina– sapphire substrates, SiC devices can operate at higher power densities than GaN or Si, with higher frequencies and better reliability. Despite this, it is difficult to measure the change in temperature gradient in nanodevices. Therefore, in this report, we use atomistic simulation methods to explore the heat transport in Si, Ge, SiC, and GeC. In this report, we focus on 4H-SiC and 4H-GeC because silicon carbide (SiC) MOSFETs (metal oxide semiconductor field effect transistor) have become mainstream products in power devices.

*Simulation Method*

We performed a first-principles calculation based on a DFT (density functional theory)-LCAO (linear combination of atomic orbitals) method as implemented in QATK (QuantumATK). Exchange–correlation function use GGA (Generalized-gradient approximation), and force tolerance set to 0.05 eV Å. To avoid underestimating the band gap in semiconductors, the DFT and HSE06 (Heyd–Scuseria–Ernzerhof) hybrid functional methods are applied. In this study, to understand the heat transport, we used the atomic non-equilibrium momentum exchange method to simulate the heat exchange process and temperature distribution inside the nanostructure, where the left unit cell and the right unit cell are kept as heat source and heat sink, respectively. The atomistic simulation setup is then relaxed by standard MD (Molecular dynamics)-NPT (Simulation of system based on constants number, pressure, and temperature). Finally, we obtained the transverse unit vector and equilibrated to the target temperature.

*Result and Discussion*

After the DFT calculation and relaxation, We used the non-equilibrium momentum exchange method to obtain thermal conductivities and the calculation results are in Fig. 1. It can be seen that the thermal conductivities of Si and in SiC are smaller than the reported literature values of 150 W K$^{-1}$ m$^{-1}$ and 450 W K$^{-1}$ m$^{-1}$, respectively. This result is expected due to the size effect. We will explain this phenomenon in the next section. SiC shows higher conductivity than Si due to the hexagonal Brillouin zone with high symmetry. The ranking of thermal conductivities are: SiC > GeC ≈ Si > Ge. The ratio of thermal conductivity kSiC/kSi ~ 2.86 is close to previous references. It is worth noting that the calculated 4H-GeC shows a 6-times higher thermal conductivity than bulk germanium. The ratio of thermal conductivity kGeC/kSi ~ 1.26 is close to that of conventional Si substrate.

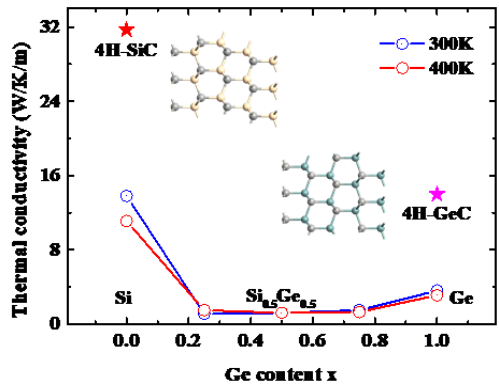

Fig.1. Thermal conductivities and summarized computing results by the non-equilibrium momentum exchange method

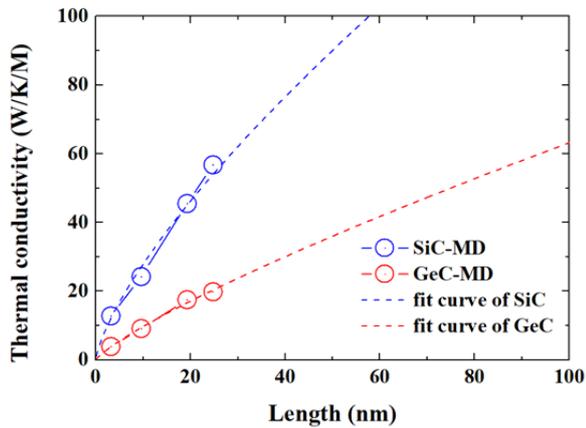

Fig.2. Thermal conductivity vs device length of 4H-SiC and 4H-GeC. Both materials are made in four lengths of 3.2 nm, 9.6 nm, 19.3 nm, and 24.8 nm. The dashed lines denote extrapolation of molecular dynamics calculations.

For the size effect, we calculated the thermal conductivity of 4H-SiC and 4H-GeC at 3.2 nm, 9.6 nm, 19.3 nm, and 24.8 nm lengths. An extrapolation was performed to obtain the thermal conductivity of the bulk, as is shown in Fig. 2. We found the 4H-SiC results compared well to other references.

## Conclusions

In this work, we show the results of first-principles simulations of the popular semiconductor channel materials silicon carbide(SiC) and germanium carbide(GeC). GeC exhibits a much higher thermal conductivity than that of bulk germanium, close to that of conventional Si substrate and high critical field than traditional Si-MOSFETs. Finally, we recommend 4H-GeC as a candidate channel material for high-performance logic applications, not only because of its wide band gap, but also because of its comparable thermal conductivity to silicon.

## Acknowledgements

This work was financially supported by the NSTC 111-2218-E-A49-016-MBK, NSTC 109-2221-E-008-093-MY3, NSTC 111-2622-8-A49-018-SB